\documentclass[12pt]{article}
\usepackage{amssymb}
\usepackage[dvips]{epsfig}
\usepackage{amsmath,amssymb,indentfirst,epsfig,bm}

\newcommand{\be}{\begin{equation}}
\newcommand{\ee}{\end{equation}}
\def\bea{\begin{eqnarray}}
\def\eea{\end{eqnarray}}

\newcommand{\bn}{\begin{eqnarray}}
\newcommand{\en}{\end{eqnarray}}

\newcommand{\no}{\noindent}

\def\bea{\begin{eqnarray}}
\def\eea{\end{eqnarray}}

\newcommand{\beq}{\begin{eqnarray}}
\newcommand{\eeq}{\end{eqnarray}}

\newcommand{\lagmzero}{\mathcal{L}^{g,m=0}(a_1)}
\newcommand{\lamsmzero}{\mathcal{L}^{\mbox{\tiny{MS}},m=0}(a_1)}
\newcommand{\la}{\mathcal{L}(a_1)}
\newcommand{\lc}{\mathcal{L}_{\mbox{\tiny{nFP}}}(c)}

\newcommand{\lcgmzero}{\mathcal{L}^{g,m=0}_{\mbox{\tiny{nFP}}}}
\newcommand{\bit}{\begin{itemize}}
\newcommand{\eit}{\end{itemize}}

\newcommand{\lag}{\mathcal{L}^g(a_1)}

\newcommand{\lcg}{\mathcal{L}^{g}_{\mbox{\tiny{nFP}}}(c)}

\newcommand{\n}{\nabla}

\newcommand{\lnfpmzero}{\mathcal{L}^{g,m=0}}
\newcommand{\lcmsmzero}{\mathcal{L}^{\mbox{\tiny{MS}},m=0}_{\mbox{\tiny{nFP}}}}
\begin{document}

\title{\textbf{A note on massless and partially massless spin-2 particles in a curved background via a nonsymmetric tensor}}
\author{H. G. M. Fortes\footnote{hemily.gomes@gmail.com}, D. Dalmazi\footnote{dalmazi@feg.unesp.br} \\
\textit{{UNESP - Campus de Guaratinguet\'a - DFQ} }\\
\textit{{Avenida Dr. Ariberto Pereira da Cunha, 333} }\\
\textit{{CEP 12516-410 - Guaratinguet\'a - SP - Brazil.} }\\}
\date{\today}
\maketitle

\begin{abstract}

\noindent In the last few years we have seen an increase interest on gravitational waves due to recent and
striking experimental results confirming Einstein's general relativity once more. From the field theory point of
view, gravity describes the propagation of self-interacting massless spin-2 particles. They can be identified
with metric perturbations about a given background metric. Since the metric is a symmetric tensor, the massless
spin-2 particles present in the Einstein-Hilbert (massless Fierz-Pauli) theory are naturally described by a
symmetric rank-2 tensor. However, this is not the only possible consistent massless spin-2 theory at linearized
level. In particular, if we add a mass term, a new one parameter $(a_1)$ family  of models ${\cal L}(a_1)$ shows
up. They consistently describe massive spin-2 particles about Einstein spaces in terms of a non-symmetric rank-2
tensor. Here we investigate the massless version of  ${\cal L}(a_1)$ in a curved background. In the case
$a_1=-1/12$ we show that the massless spin-2 particles consistently propagate, at linearized level, in maximally
symmetric spaces. A similar result is obtained otherwise $(a_1 \ne -1/12)$ where we have a non-symmetric
scalar-tensor massless model.  The case of partially massless non-symmetric models is also investigated.

\end{abstract}

\newpage

\section{Introduction}

The recent increase of the studies on massive spin-2 particles \cite{hinter,drham} is partially due to the fact
they can represent massive gravitons which may offer an alternative explanation for the accelerated expansion of
the universe since they lead to a weaker gravitational interaction at large distances \cite{super1,super2}.
Notice however that there are very low experimental upper bounds on the graviton mass, for instance from the
LIGO experiment of detection of gravitational waves one has $10^{-22} eV$, see \cite{gw,gw2}.

Another motivation is the quite recent overcome of  historical theoretical obstacles in the description of
massive gravitons, like the vDVZ mass discontinuity \cite{vdv,zak} and the existence of ghosts in the nonlinear
theory \cite{db}. They have been solved by the addition of fine-tuned nonlinear self-interaction terms for the
graviton \cite{drgt,hr}. In 2015, it was obtained from the dRGT models of \cite{drgt} a linear covariant theory
consistent with the description of massive gravitons propagating on arbitrary backgrounds \cite{new1,BDvS}.
Thus, recovering previous perturbative results of \cite{Pershin1,Pershin2}.

All those studies of massive spin-2 particles have considered the Fierz-Pauli (FP) theory \cite{FierzPauli} as
their starting point. The FP description is based on a symmetric and traceful rank-2 tensor
$h_{\mu\nu}=h_{\nu\mu}$ which propagates 5 degrees of freedom (d.o.f) in $D=4$. It can be seen as the metric
fluctuation about a background metric $g_{\mu\nu}^{(0)}$.

In \cite{Dalmazi2} another family of models $\la$, where $a_1$ is an arbitrary real constant, has been suggested
which describes massive spin-2 particles via a  nonsymmetric rank-2 tensor $e_{\mu\nu}\ne e_{\nu\mu}$ in flat
spaces. We have \cite{Fortes} coupled a background gravitational field to $\la$ by including also nonminimal
terms and have looked for curved space generalizations of the tensor, vector and scalar constraints which are
necessary in order to get rid of nonphysical degrees of freedom. We require that the coefficients of the
nonmininal terms be analytic functions of $m^2$. Such restriction leads us to constraint the gravitational
background to Einstein spaces.

Regarding the massless case, some authors consider the linearized Einstein-Hilbert theory (massless Fierz-Pauli)
as the only possible description of massless spin-2 particles via a rank-2 tensor, see the earlier work
\cite{pun}, except eventually for the WTDiff (Weyl and transverse diffeomorphism) invariant theory, see e.g.
\cite{abgv}.  If the massless Fierz-Pauli theory is embedded in a curved background, the required vector
symmetry implies that the background must be of Einstein type, i.e., $R_{\mu\nu}=R \, g_{\mu\nu} /D \, $, which
is the vacuum solution for Einstein equations with cosmological constant, see \cite{Henn}. Neither the addition
of non-minimal higher derivative terms nor allowing non-analytic terms in the cosmological constant change this
result. The case of the WTDIFF theory also requires Eintein spaces \cite{Hinter2}.

In the present work, we look for the massless version of $\la$ coupled to a background, providing another
possible description for massless spin-2 particles besides the massless FP model. We can compare our conclusions
with those of \cite{Henn}.

Still, when we deal with curved spaces, there is a different situation which deserves a special attention. On
maximally symmetric spaces, there is a specific value for the curvature constant $R$ in terms of $m^2$ which
allows us to have a scalar gauge symmetry, even with $m\neq0$. Consequently, we have a theory describing a
massive spin-2 particle with 4 degrees of freedom, instead of $5=2s+1$ for $D=4$. This kind of theory has been
intensively studied in massive gravity and it is called partially massless theories \cite{dn}. Thus, in the
present work, we seek the partially massless theories corresponding to the $\lag$ models.

\section{Spin-2 particles in curved spaces}

\subsection{Fierz-Pauli action}
\label{ssFPm0}

The linear action for massive spin-2 particles propagating on a curved background $g_{\mu\nu}$ is usually
described by the linearized Einstein-Hilbert action plus the Fierz-Pauli mass term and an extra term
proportional to the scalar curvature\footnote{Throughout this work we use $\eta_{\mu\nu}=(-,+,+,+)$}:
\begin{eqnarray}
S=\int d^4 x\sqrt{-g}\biggl[ -\frac{1}{2}\n_\alpha h_{\mu\nu}\n^\alpha h^{\mu\nu}+\n_\alpha h_{\mu\nu}\n^\nu h^{\mu\alpha} -\n_\mu h \n_\nu h^{\mu\nu}+\frac{1}{2}\n_\mu h \n^\mu h+\nonumber \\
-\frac{1}{2}m^2(h_{\mu\nu}h^{\mu\nu}-h^2)+\frac{R}{4}\biggl(h^{\mu\nu}h_{\mu\nu}-\frac{1}{2}h^2\biggr)\biggr]
\label{FPaction}
\end{eqnarray}

\no where $h_{\mu\nu}$ is a symmetric tensor ($h_{\mu\nu} = h_{\nu\mu}$). The covariant derivatives are
calculated with respect to a background metric $g_{\mu\nu}^{(0)}$. In the flat space,
$g_{\mu\nu}^{(0)}=\eta_{\mu\nu}$, the theory (1) becomes the usual Fierz-Pauli theory whose
massless version is the linearized Einstein-Hilbert model $(\sqrt{-g}R)_{hh}$ where $g_{\mu\nu} =
\eta_{\mu\nu} + h_{\mu\nu}$.

In order to have consistency, it is necessary to obtain all the curved space Fierz-Pauli constraints,
\begin{eqnarray}
\n^\mu h_{\mu\nu}&=&0\label{vector}\\
h&=&0 \ ,\label{scalar}
\end{eqnarray}
which is achieved only in Einstein background spaces, see, e.g., \cite{hinter} and references therein,
\begin{eqnarray}
R_{\mu\nu}=\frac{R}{4}g_{\mu\nu}\ . \label{es4}
\end{eqnarray}
When we seek for the scalar constraint (\ref{scalar}), the expression below comes up when we combine second
derivatives $\nabla_{\mu}\nabla_{\nu}E^{\mu\nu}$ and the trace $g_{\mu\nu}E^{\mu\nu}$ of the equations of motion
$E^{\mu\nu}=0$:
\begin{eqnarray}
\biggl( 3m^2-\frac{R}{2}\biggr) h=0 \ . \label{m2R}
\end{eqnarray}

Thus, depending on the value of $m^2$, we have models with different particle contents. Let us see the main
results:

{\bf i)} As far as $m^2 \neq R/6$ and $m\neq 0$, besides the four constraints (\ref{vector}) we obtain from
(\ref{m2R}) the desired scalar constraint $h=0$. In this case, we are left with 5 propagating d.o.f. as expected
for a massive spin-2 model.

{\bf ii)} If $m^2=R/6$, we loose the scalar constraint\footnote{The case $m^2=R/6$ corresponds to the so-caled
Higuchi bound \cite{pm1}.} (\ref{scalar}). Instead, the action has a scalar gauge symmetry
\begin{eqnarray}
\delta h_{\mu\nu}=\n_\mu \n_\nu \phi+\frac{m^2}{2}\phi g_{\mu\nu} \ \ .
\end{eqnarray}
A scalar gauge symmetry removes two degrees of freedom in contrast to a scalar constraint which removes only
one. As a result, the model propagates 4 d.o.f. instead of 5 which is known as a ``partially massless'' theory
\cite{dn} and it will be discussed in the subsection \ref{pmtheories}.

{\bf iii)} If $m=0$, there is neither vector nor scalar constraint but conversely the action acquires the vector
gauge symmetry
\begin{eqnarray}
\delta h_{\mu\nu}=\n_\mu \xi_\nu +\n_\nu \xi_\mu\label{eq13}
\end{eqnarray}
which is the linearized diffeomorphism symmetry of general relativity. We are left with 2 degrees of freedom,
describing in fact a massless spin-2 particle. Furthermore, at $m=0$ the action (\ref{FPaction}) coincides with
the linearized version of
\begin{eqnarray}
S_\Lambda=
\int d^4 x \, \sqrt{-g}\,(R-2\Lambda)
\end{eqnarray}
around a curved background $g_{\mu\nu}=g^{\mbox{\tiny{(0)}}}_{\mu\nu}+h_{\mu\nu}$ of Einstein kind with
$R=4\Lambda$ which seems to indicate that any massless spin-2 particle must be identified with the graviton
\cite{Henn} as we have mentioned before.

\subsection{$\la$ models and their massless versions}

In \cite{Annals} a family of second order Lagrangians $\la$ has been presented in the flat space and in
arbitrary dimensions $D\geq 3$, but here we focus in $D=4$. It describes massive spin-2 particles via a
nonsymmetric rank-2 tensor $e_{\mu\nu}\neq e_{\nu\mu}$. In \cite{Fortes} the $\la$ models have been coupled to a
curved background. In order to find massive theories on curved spaces we minimally couple the corresponding flat
space action and then add curvature terms in such a way to obtain the necessary constraints and achieve the
correct number of degrees of freedom,

\begin{eqnarray}
\lag &=& -\frac{1}{4}\nabla^\mu e^{\alpha\beta}\,\nabla_\mu e_{\alpha\beta}-\frac{1}{4}\nabla^\mu e^{\alpha\beta}\,\nabla_\mu e_{\beta\alpha}+a_1\nabla^\alpha e_{\alpha\beta}\,\nabla_\mu e^{\mu\beta}+\frac{1}{2}\nabla^\alpha e_{\alpha\beta}\,\nabla_{\mu}e^{\beta\mu}+\nonumber \\
&\,&+\frac{1}{4}\nabla^\alpha e_{\beta\alpha}\,\nabla_{\mu}e^{\beta\mu}
+\biggl(a_1+\frac{1}{4}\biggr)\nabla^\mu e\,\nabla_\mu e-\biggl(a_1+\frac{1}{4}\biggr)\nabla^\mu e\,\nabla^\alpha \, e_{\alpha\mu}+\nonumber \\
&\,& -\biggl(a_1+\frac{1}{4}\biggr)\nabla^\mu e \,\nabla^{\alpha}e_{\mu\alpha}+f_1\,R\,e^{\alpha\beta}\,e_{\alpha\beta}+f_2\,R\,e^2+f_3\,R_{\alpha\beta\mu\nu}\,e^{\alpha\mu}\,e^{\beta\nu}+\nonumber \\
&\,& +f_4\,R_{\alpha\beta}\,e^{\alpha\mu}\,{e^{\beta}}_\mu +f_5\,R_{\alpha\beta}\,e^{\alpha\beta}\,e+f_6\,R_{\alpha\beta\mu\nu}\,e^{\alpha\beta}\,e^{\mu\nu}+f_7\,R_{\alpha\beta}\,e^{\alpha\mu}\,{e_\mu}^\beta+\nonumber \\
&\,& +f_8\,R\,e^{\alpha\beta}\,e_{\beta\alpha}+f_9\,R_{\alpha\beta}\,e^{\mu\alpha}\,{e_{\mu}}^\beta -
\frac{m^2}2(e_{\mu\nu}e^{\nu\mu} - e^2) \label{La1gmzero}
\end{eqnarray}

\no The constant $a_1$ is a real number and $e=g^{\mu\nu}e_{\mu\nu}$. The coefficients $f_j$, $j=1\cdots 9$,
are partially fixed \cite{Fortes} by requiring that the curved space FP constraints are satisfied:

\begin{eqnarray}
\n^\mu e_{\mu\nu}=0=e_{[\mu,\nu]}=0=g^{\mu\nu}e_{\mu\nu}\label{cfp2} \end{eqnarray}

\no In the flat space we recover the theory $\mathcal{L}(a_1)$ of \cite{Dalmazi2} which describes massive spin-2
particles and whose massless part is invariant under

\begin{eqnarray}
\delta e_{\mu\nu}=\partial_\nu \xi_\mu + \partial^\alpha \Lambda_{[\alpha\mu\nu]} \ , \label{sym1}
\end{eqnarray}

\no with $\Lambda_{[\alpha\mu\nu]}$ a fully antisymmetric tensor. It is interesting to split the discussion into
three cases. At $a_1=1/4$ we recover the FP model, since the antisymmetric components
$(e_{\mu\nu}-e_{\nu\mu})/2$ decouple due to the enlargement of the massless symmetries (\ref{sym1}) by
antisymmetric shifts $\delta e_{\mu\nu}=\Lambda_{\mu\nu}=-\Lambda_{\nu\mu}$. At $a_1 = - 1/12$ the massless
symmetries (\ref{sym1}) are augmented by Weyl transformations $\delta e_{\mu\nu}=\eta_{\mu\nu}\phi$. Finally, at
$a_1 \ne 1/4$ and $a_1 \ne -1/12$, the particle content of $\mathcal{L}^{m=0}(a_1)$ consists of massless spin-2
particles plus massless spin-0 particles. The massless spin-0 particle is physical if $a_1 > \frac{1}{4}$ or
$a_1 < -\frac{1}{12}$ and disappears at $a_1 \ne 1/4$ or $a_1 \ne -1/12$ whereas the spin-2 particle is always
physical.

Regarding massless theories on curved spaces, we now require invariance under gauge symmetries. As in the flat
case we have three cases: $a_1=1/4$, $a_1=-1/12$ and $a_1\neq 1/4,-1/12$. Since we recover the known FP theory
at $a_1=1/4$, we start with $a_1=-1/12$ where we slightly change the notation from $f_j$ to $d_j$. Due to the
Weyl symmetry in the massless sector there is no need anymore of the FP fine tuning of the mass term and the
model is called a non-Fierz-Pauli one with any arbitrary constant $c$ in the mass term, see (\ref{lcg}).

\subsubsection{$\lc$ model ($a_1=-1/12$)}
The generalization of the massive theory $\lc$ to curved spaces was first suggested in \cite{Fortes} and the
main results are summarized in the equations (\ref{lcg})-(\ref{mtil2}) below. The most general Lagrangian
coupled to an Einstein background and quadratic in derivatives is given by \bea
{\mathcal{L}}_{\mbox{\tiny{nFP}}}^g(c)&=&\sqrt{-g}\biggl[-\frac{1}{4}\n^\mu e^{\alpha\beta}\n_\mu e_{\alpha\beta}-\frac{1}{4}\n^\mu e^{\alpha\beta}\n_\mu e_{\beta\alpha}-\frac{1}{12}\n^\alpha e_{\alpha\beta}\n_\lambda e^{\lambda\beta}+\frac{1}{2}\n^\alpha e_{\alpha\beta}\n_\lambda e^{\beta\lambda}+\nonumber \\
&\,&+\frac{1}{4}\n^\alpha e_{\beta\alpha} \n_\lambda e^{\beta\lambda}+\frac{1}{6}\n^\mu
\n_\mu e-\frac{1}{3}\n^\alpha e_{\alpha\beta}\n^\beta e-
\frac{m^2}{2}(e_{\alpha\beta}e^{\beta\alpha}+c\, e^2)+\nonumber \\
&\,&+d_1\, R\,e^{\alpha\beta}\,e_{\alpha\beta}+d_2\,R\,e^2+d_3\, R_{\alpha\beta\mu\nu}\,e^{\alpha\mu}\,e^{\beta\nu}+d_4\,R_{\alpha\beta}\,e^{\alpha\mu}\,{e^{\beta}}_\mu+d_5\, R_{\alpha\beta}\,e^{\alpha\beta}\,e+\nonumber \\
&\,&+d_6 \,R_{\alpha\beta\mu\nu}\,e^{\alpha\beta}\,e^{\mu\nu}+d_7\,
R_{\alpha\beta}\,e^{\alpha\mu}\,{e_\mu}^\beta+d_8\,R \,e^{\alpha\beta}\, e_{\beta\alpha}+d_9
\,R_{\alpha\beta}\,e^{\mu\alpha}\,{e_\mu}^\beta\biggr] \label{lcg} \eea where $d_j$'s are free parameters {\it a
priori}. It is necessary that the model presents the correct number of degrees of freedom. If we require that
$d_j$ are all analytic functions of $m^2$, in order to satisfy the  FP constraints (\ref{cfp2}) it is necessary
\cite{Henn} to restrict the background to Einstein spaces (\ref{es4}) and fix three parameters,
\begin{eqnarray}
{d_1}+\frac{d_4}{4}+\frac{d_9}{4}=0\ ,\ d_3=1 \ , \ d_6=-\frac{1}{2} \label{soldj}
\end{eqnarray}

The equations of motion become\footnote{Notice that the equations of motion are not exactly the Klein-Gordon
ones since they present an additional term with the Riemann curvature. By considering that the transverse
condition must be satisfied, i.e., $\nabla^\mu h_{\mu\nu}=0$, the presence of such term is required. Otherwise,
there would be an inconsistency in the calculation of the commutator $[\n^\mu,\square -m^2]h_{\mu\nu}$, which is
non-null.}
\begin{eqnarray}
E_{\rho\sigma}=(\square -\tilde{m}^2)\, e_{\rho\sigma}+2R_{\rho\alpha\sigma\beta}\ e^{\alpha\beta}=0
\end{eqnarray}
where
\begin{eqnarray}
\tilde{m}^2\equiv m^2-\biggl(2{d}_8+\frac{d_7}{2}\biggr)R\label{mtil2}\ .
\end{eqnarray}

Let us now consider the massless version of (\ref{lcg}). The gauge symmetries of the flat case are given in
(\ref{sym1}) plus Weyl transformations. Now we expect
\begin{eqnarray}
\delta e_{\mu\nu}= g_{\mu\nu} \phi +\n_\nu \xi_\mu + \n^\alpha \Lambda_{[\alpha\mu\nu]}.\label{transflc}
\end{eqnarray}
By calculating the variation of the action $S^{g,m=0}_{\mbox{\tiny{nFP}}}$ under (\ref{transflc}), we obtain
(under the integral):
\begin{eqnarray}
\delta\lcgmzero &=& \sqrt{-g}\Biggl\{\phi\biggl[(2d_1+8d_2+d_5+2d_8)R\, e+2(d_3+d_4+2d_5+d_7+d_9)R_{\rho\sigma}e^{\rho\sigma}\biggr]\nonumber \\
&\,& +\n_\nu \xi_\mu\biggl[\biggl(\frac{1}{6}+d_7\biggr){R^\mu}_\alpha \, e^{\nu\alpha}+\biggl(\frac{1}{2}+d_7\biggr){R^\nu}_\alpha \, e^{\alpha\mu}+2(-1+d_3)R^{\mu\beta\nu\rho}\, e_{\beta\rho}\nonumber \\
&\,& +(1+2d_6)R^{\alpha\beta\mu\nu}\, e_{\alpha\beta}+\biggl(-\frac{1}{2}+2d_4\biggr){R^\mu}_\beta \, e^{\beta\nu}+\biggl(\frac{1}{2}+2d_9\biggr){R^\nu}_\beta \, e^{\mu\beta}+\nonumber \\
&\,&+\biggl(\frac{1}{3}+d_5\biggr)R^{\nu\mu}\, e +2d_1R\, e^{\mu\nu}+2d_8 R e^{\nu\mu}\biggr]+\n^\mu\, \xi_\mu\biggl[2d_2R\, e +d_5 R^{\alpha\beta}\, e_{\alpha \beta}\biggr] \nonumber \\
&\,& +\xi_\mu\biggl[-\frac{1}{3}\, e_{\lambda\nu}\, \n^\lambda R^{\mu\nu}- e_{\lambda\nu}\, \n^\nu R^{\mu\lambda}+ e_{\alpha\lambda}\n^\mu R^{\lambda\alpha} +\frac{1}{6}\, e\, \n_\mu R\biggr]+\nonumber \\
&\,& +\biggl[2R(d_1-d_8)e^{\mu\nu}+(2d_4-d_7){R^\mu}_\beta \, e^{\beta\nu}+(2d_9-d_7){R^\nu}_\beta \, e^{\mu\beta}\biggr]\n^\alpha \Lambda_{[\alpha\mu\nu]}\Biggr\}\ .\nonumber \\
\end{eqnarray}
We have not been able to get $\delta S^{g,m=0}_{\mbox{\tiny{nFP}}}=0$ by choosing the coefficients $d_j$'s. This
can be noticed if we look specifically at the coefficients $(\frac{1}{6}+d_7)$ and $(\frac{1}{2}+d_7)$ which can
not be canceled simultanously. That is why we are going to restrict the background to the Einstein spaces
similarly to what happened in the massive case \cite{Fortes}.

Therefore let us reconsider the $\lcgmzero$ model coupled to Einstein spaces (\ref{es4}). Now we have five free
parameters:
\begin{eqnarray}
\lnfpmzero &=&-\frac{1}{4}\n^\mu e^{\alpha\beta}\n_\mu e_{\alpha\beta}-\frac{1}{4}\n^\mu e^{\alpha\beta}\n_\mu e_{\beta\alpha}-\frac{1}{12}\n^\alpha e_{\alpha\beta}\n_\lambda e^{\lambda\beta}+\frac{1}{2}\n^\alpha e_{\alpha\beta}\n_\lambda e^{\beta\lambda}+\nonumber \\
&\,&+\frac{1}{4}\n^\alpha e_{\beta\alpha} \n_\lambda e^{\beta\lambda}+\frac{1}{6}\n^\mu \n_\mu e-\frac{1}{3}\n^\alpha e_{\alpha\beta}\n^\beta e+\tilde{d_1}\, R\,e^{\alpha\beta}\,e_{\alpha\beta}+\nonumber \\
&\,&+\tilde{d_2}\,R\,e^2+d_3\, R_{\alpha\beta\mu\nu}\,e^{\alpha\mu}\,e^{\beta\nu}+d_6
\,R_{\alpha\beta\mu\nu}\,e^{\alpha\beta}\,e^{\mu\nu}+\tilde{d_8}\,R \,e^{\alpha\beta}\, e_{\beta\alpha}
\label{lcgm0es}
\end{eqnarray}
where we have defined
\begin{eqnarray}
\tilde{d_1}&\doteq & d_1+\frac{d_4}{4}+\frac{d_9}{4}\label{d1til}\\
\tilde{d_2}& \doteq & d_2+\frac{d_5}{4}\label{d2til}\\
\tilde{d_8}& \doteq & d_8+\frac{d_7}{4} \ . \label{d8til}
\end{eqnarray}

Under the integral we have:
\begin{eqnarray}
&\,& \delta\lcgmzero =\sqrt{-g}\Biggl\{\phi\biggl(2\tilde{d_1}+8\tilde{d_2}+\frac{d_3}{2}+2\tilde{d_8}\biggr)R\, e +\biggl(\frac{1}{12}+2\tilde{d_2}\biggr)R\, e\, \n^\mu \xi_\mu +\nonumber \\
&\,&
+\biggl[(-1+2d_3+2d_6)R^{\mu\beta\nu\alpha}e_{\beta\alpha}-(1+2d_6)R^{\alpha\nu\beta\mu}e_{\alpha\beta}+\biggl(\frac{1}{6}+2\tilde{d_8}\biggr)R\,
e^{\nu\mu}
+2\tilde{d_1}R\, e^{\mu\nu}\biggr]\n_\nu \xi_\mu \nonumber \\
&\,& +\biggl[(d_3+2d_6)R^{\mu\beta\nu\lambda}e_{\beta\lambda}+(2\tilde{d_1}-2\tilde{d_8})R\,
e^{\mu\nu}\biggr]\n^\alpha \Lambda_{[\alpha\mu\nu]}\Biggr\}\label{variacaoLc}
\end{eqnarray}

Therefore, in order to have Weyl invariance we need:
\begin{eqnarray}
2\tilde{d_1}+8\tilde{d_2}+\frac{d_3}{2}+2\tilde{d_8}=0\ .\label{eq56}
\end{eqnarray}
On the other hand, invariance under the vector transformation $\delta e^{(2)}_{\mu\nu}= \n_\nu\xi_\mu$ needs:
\begin{eqnarray}
-1+2d_3+2d_6=0\\
1+2d_6=0\\
\frac{1}{6}+2\tilde{d_8}=0\label{d8t1}\\
\tilde{d_1}=0\label{d1t}\\
\frac{1}{12}+2\tilde{d_2}=0\label{d2t}
\end{eqnarray}
Finally, in order to get invariance under $\delta e^{(3)}_{\mu\nu}= \n^\alpha \Lambda_{[\alpha\mu\nu]}$, we
demand:
\begin{eqnarray}
d_3+2d_6=0\label{d3d6}\\
\tilde{d_1}-\tilde{d_8}=0\label{d8t2}
\end{eqnarray}
We see from the equations (\ref{d8t1}), (\ref{d1t}) and (\ref{d8t2}) that there is no solution which makes the
Lagrangian invariant under the transformations $\delta e^{(2)}_{\mu\nu}= \n_\nu\xi_\mu$ and $\delta
e^{(3)}_{\mu\nu}= \n^\alpha \Lambda_{[\alpha\mu\nu]}$ simultaneously. Therefore, it is not possible to obtain in
this case a consistent model for massless spin-2 particles propagating even on Einsteins spaces. From this point
of view, regarding the massless case, the model with a symmetric field $h_{\mu\nu}=h_{\nu\mu}$ given in
(\ref{FPaction}) is more flexible than $\lcg$. Still, we can identify two cases with partial symmetries:
\begin{itemize}
\item {\textbf{Scalar and vector symmetries}}

It is possible to find a unique solution for the reduced system of equations (\ref{eq56})-(\ref{d2t}):
\begin{eqnarray}
\tilde{d_1}=0, \ \tilde{d_2}=-\frac{1}{24}, \ d_3=1, \ d_6=-\frac{1}{2}, \ \tilde{d_8}=-\frac{1}{12}
\end{eqnarray}
In this case, we have a model invariant under the following gauge transformation:
\begin{eqnarray}
\delta e_{\mu\nu}=g_{\mu\nu} \phi +\n_\nu \xi_\mu \ .
\end{eqnarray}

\item {\textbf{Scalar and tensor symmetries}}

Analogously, from (\ref{eq56}), (\ref{d3d6}) and (\ref{d8t2}), we have a model invariant under the gauge
transformation below:
\begin{eqnarray}
\delta e_{\mu\nu}=g_{\mu\nu} \phi + \n^\alpha \Lambda_{[\alpha\mu\nu]}\ ,
\end{eqnarray}
where we need
\begin{eqnarray}
\tilde{d_1}=\tilde{d_8}, \ d_3=-2d_6, \ \tilde{d_2}=\frac{d_6}{8}-\frac{\tilde{d_8}}{2}\ .
\end{eqnarray}
\end{itemize}

Now, we consider the model $\lcgmzero$ coupled to maximally symmetric (MS) spaces which are spaces whose
Riemmann tensor is given by
\begin{eqnarray}
R_{\alpha\beta\rho\sigma}=\frac{R}{12}(g_{\alpha\rho}g_{\beta\sigma}-g_{\alpha\sigma}g_{\beta\rho}) \
.\label{RMS}
\end{eqnarray}
The variation of the Lagrangian (\ref{variacaoLc}) can be rewrite as:
\begin{eqnarray}
\delta\lcmsmzero &=&\sqrt{-g}R\Biggl\{\phi\biggl(2\tilde{d_1}+8\tilde{d_2}+\frac{d_3}{2}+2\tilde{d_8}\biggr) e +\nonumber \\
&\,& +\biggl(-\frac{1}{12}+\frac{d_3}{6}+2\tilde{d_2}\biggr) e\, \n^\mu \xi_\mu+\biggl(\frac{1}{4}-\frac{d_3}{6}-\frac{d_6}{6}+2\tilde{d_8}\biggr) e^{\nu\mu}\, \n_\nu\xi_\mu +\nonumber \\
&\,& +\biggl(\frac{1}{12}+\frac{d_6}{6}+2\tilde{d_1}\biggr)e^{\mu\nu}\, \n_\nu \xi_\mu
+\biggl(\frac{d_3}{12}+\frac{d_6}{6}+2\tilde{d_1}-2\tilde{d_8}\biggr)e^{\mu\nu}\,
\n^\alpha\Lambda_{[\alpha\mu\nu]}
\Biggr\}\nonumber \\
\label{variacaoLcms}
\end{eqnarray}
In order to get $\delta\lcmsmzero=0$, each coefficient of the expression above must be null, which means that we
need to solve the system below:
\begin{eqnarray}
2\tilde{d_1}+8\tilde{d_2}+\frac{d_3}{2}+2\tilde{d_8}=0 \label{eq23}\\
-\frac{1}{12}+2\tilde{d_2}+\frac{d_3}{6}=0\\
\frac{1}{4}-\frac{d_3}{6}-\frac{d_6}{6}+2\tilde{d_8}=0\\
\frac{1}{12}+2\tilde{d_1}+\frac{d_6}{6}=0\\
2\tilde{d_1}+\frac{d_3}{12}+\frac{d_6}{6}-2\tilde{d_8}=0\label{eq27}
\end{eqnarray}
and the solution found is
\begin{eqnarray}
\tilde{d_1}=-\frac{1}{12}+\tilde{d_8}, \, \tilde{d_2}=-\frac{1}{24}-2\, \tilde{d_8}, \, d_3=1+24\, \tilde{d_8},
\, d_6=\frac{1}{2}-12\, \tilde{d_8}\label{sol528}
\end{eqnarray}
where $\tilde{d_8}$ is still arbitrary because only four of the five equations (\ref{eq23})-(\ref{eq27}) are
independent. The existence of such solution means that $\lcgmzero$ in maximally symmetric spaces has symmetry
under the complete set of transformations (\ref{transflc}).

When we substitute (\ref{RMS}) and the solution (\ref{sol528}) in (\ref{lcg}) at $m=0$, the original Lagrangian
becomes:
\begin{eqnarray}
\lcmsmzero &=&-\frac{1}{4}\n^\mu e^{\alpha\beta}\n_\mu e_{\alpha\beta}-\frac{1}{4}\n^\mu e^{\alpha\beta}\n_\mu e_{\beta\alpha}-\frac{1}{12}\n^\alpha e_{\alpha\beta}\n_\lambda e^{\lambda\beta}+\frac{1}{2}\n^\alpha e_{\alpha\beta}\n_\lambda e^{\beta\lambda}+\nonumber \\
&\,&+\frac{1}{4}\n^\alpha e_{\beta\alpha} \n_\lambda e^{\beta\lambda}+\frac{1}{6}\n^\mu \n_\mu e-\frac{1}{3}\n^\alpha e_{\alpha\beta}\n^\beta e-\frac{1}{24}\, R\,e^{\alpha\beta}\,e_{\alpha\beta}+\nonumber \\
&\,&+\frac{1}{24}\,R\,e^2-\frac{1}{8}\,R \,e^{\alpha\beta}\, e_{\beta\alpha}\label{lcms2}
\end{eqnarray}
where $\tilde{d_8}$ ends up being eliminated from the coefficients. Therefore, we have a unique model consistent
with the description of massless spin-2 particles propagating in maximally symmetric spaces. In addition, the
Lagrangian (\ref{lcms2}) describes the massless limit of the massive model $\lcg$ given in (\ref{lcg}) with
$\tilde{d_2}=-1/24$ and $\tilde{d_8}=-1/12$. It allows us to conclude that such subcase of the massive model
$\lcg$ has a consistent massless limit, at least in maximally symmetric spaces.

\subsubsection{$\la$ with $a_1\neq (1/4,-1/12)$}

Similarly to the previous subsection, we need the massless version of the local symmetries on curved spaces
given in (\ref{sym1}) in order to obtain a consistent massless version of $\lag$:
\begin{eqnarray}
\delta e_{\mu\nu}=\n_\nu \xi_\mu +\n^\alpha \Lambda_{[\alpha\mu\nu]}\ . \label{symla}
\end{eqnarray}

Let us start from the most general Lagrangian (9) with $m=0$ where
 $f_j$ are arbitrary constants for now. The variation leads (under the integral) to
\begin{eqnarray}
\delta\lagmzero &=& \sqrt{-g}\Biggl\{\n_\nu \xi_\mu\biggl[\biggl(-2a_1+f_7\biggr){R^\mu}_\alpha \, e^{\nu\alpha}+\biggl(\frac{1}{2}+f_7\biggr){R_\alpha}^\nu \, e^{\alpha\mu}+\nonumber \\
&\,& +(-1+2f_3+2f_6)R^{\mu\beta\nu\rho}\, e_{\beta\rho}+(1+2f_6)R^{\nu\alpha\beta\mu}\, e_{\alpha\beta}+\nonumber \\
&\,&+\biggl(-\frac{1}{2}+2f_4\biggr){R^\mu}_\beta \, e^{\beta\nu}+\biggl(\frac{1}{2}+2f_9\biggr){R^\nu}_\beta \, e^{\mu\beta}+\nonumber \\
&\,&+\biggl(\frac{1}{2}+2a_1+f_5\biggr)R^{\nu\mu}\, e +2f_1R\, e^{\mu\nu}+2f_8 R e^{\nu\mu}\biggr]+\nonumber \\
&\,&+\n^\mu\, \xi_\mu\biggl[2f_2R\, e +f_5 R^{\alpha\beta}\, e_{\alpha \beta}\biggr] \nonumber \\
&\,& +\xi_\mu\biggl[-2a_1\, e_{\lambda\nu}\, \n_\alpha R^{\alpha\mu\lambda\nu}-\frac{1}{2} e_{\alpha\lambda}\, \n_\nu R^{\mu\alpha\nu\lambda}-\frac{1}{2} e_{\alpha\lambda}\n_\nu R^{\mu\lambda\nu\alpha}+\nonumber \\
&\,&+\biggl(2a_1+\frac{1}{2}\biggr)( e\, \n_\nu R^{\mu\nu}-\, e_{\alpha\beta}\n^\beta R^{\mu\alpha})\biggr]\nonumber \\
&\,& +\biggl[2R(f_1-f_8)e^{\mu\nu}+(2f_4-f_7){R^\mu}_\beta \, e^{\beta\nu}+(2f_9-f_7){R^\nu}_\beta \, e^{\mu\beta}\biggr]\n^\alpha \Lambda_{[\alpha\mu\nu]}\Biggr\}\ .\nonumber \\
\end{eqnarray}
Once again we have not been able to find a solution for the $f_j$ in such way that $\delta \lagmzero=0$ on
arbitrary backgrounds. Therefore, again we consider $R_{\mu\nu}=\frac{1}{4}R \, g_{\mu\nu}$ and rewrite the
variation above:
\begin{eqnarray}
\delta\lagmzero &=& \sqrt{-g}\Biggl\{\n_\nu \xi_\mu\biggl[(-1+2f_3+2f_6)R^{\mu\beta\nu\rho}\, e_{\beta\rho}+(1+2f_6)R^{\nu\alpha\beta\mu}\, e_{\alpha\beta}+\nonumber \\
&\,&+2\tilde{f_1}R\, e^{\mu\nu}+\biggl(\frac{1}{8}-\frac{a_1}{2}+2\tilde{f_8}\biggr) R e^{\nu\mu}\biggr]+\n^\mu\, \xi_\mu\biggl[\frac{1}{8}+\frac{a_1}{2}+2\tilde{f_2}\biggr]R\, e \nonumber \\
&\,&
+\biggl[(f_3+2f_6)R^{\mu\beta\nu\lambda}e_{\beta\lambda}+2(\tilde{f_1}-\tilde{f_8})Re^{\mu\nu}\biggr]\n^\alpha\Lambda_{[\alpha\mu\nu]}\Biggr\}\label{var}
\end{eqnarray}
where we have defined
\begin{eqnarray}
\tilde{f_1}&=&f_1+\frac{f_4}{4}+\frac{f_9}{4} \\
\tilde{f_2}&=&f_2+\frac{f_5}{4} \\
\tilde{f_8}&=&f_8+\frac{f_7}{4} \ .
\end{eqnarray}
In order to obtain $\delta\lagmzero=0$, we need to find  a solution of the equations below:
\begin{eqnarray}
-1+2f_3+2f_6=0\label{538}\\
1+2f_6=0\label{539}\\
\tilde{f_1}=0\label{f1t}\\
\frac{1}{8}-\frac{a_1}{2}+2\tilde{f_8}=0\label{f8t}\\
\frac{1}{8}+\frac{a_1}{2}+2\tilde{f_2}=0\label{542}\\
f_3+2f_6=0\label{543}\\
\tilde{f_1}-\tilde{f_8}=0\label{f1tf8t}
\end{eqnarray}
However, the equations (\ref{f1t}), (\ref{f8t}) and (\ref{f1tf8t}) lead us to the Fierz-Pauli massless model:
$a_1=1/4$. Thus, it is not possible to obtain a massless model for $\lag$ ($a_1\neq 1/4$) on Einstein spaces
symmetric under (\ref{symla}).

On the other hand, we have models with vector and tensor symmetries separately:
\begin{itemize}
\item{\textbf{Vector Symmetry}}

From the equations (\ref{538})-(\ref{542}), we have the following solution:
\begin{eqnarray}
\tilde{f_1}=0, \, f_6=-\frac{1}{2},\, f_3=1, \, \tilde{f_8}=\frac{1}{4}\biggl(a_1-\frac{1}{4}\biggr), \,
\tilde{f_2}=-\frac{1}{4}\biggl(a_1+\frac{1}{4}\biggr)
\end{eqnarray}
In this specific case, the Lagrangian is invariant under the transformation $\delta^{(1)} e_{\mu\nu}=\n_\nu
\xi_\mu$.
%
%
\item{\textbf{Tensor Symmetry}}

Similarly, if we choose the parameters in such way that the equations (\ref{543}) and (\ref{f1tf8t}) are
satisfied, i.e,
\begin{eqnarray}
f_3=-2f_6, \, \tilde{f_1}=\tilde{f_8}\ ,
\end{eqnarray}
the Lagrangian becomes invariant under $\delta e_{\mu\nu}=\n^\alpha \Lambda_{[\alpha\mu\nu]}$.
\end{itemize}
Now, considering maximally symmetric spaces, the variation of the Lagrangian $\lagmzero$ can be written (under
integral) as:
\begin{eqnarray}
\delta\lamsmzero &=&\sqrt{-g}R\Biggl\{\biggl(-\frac{1}{24}+\frac{a_1}{2}+\frac{f_3}{6}+2\tilde{f_2}\biggr) e\, \n^\mu \xi_\mu+\nonumber \\
&\,&+\biggl(\frac{5}{24}-\frac{a_1}{2}-\frac{f_3}{6}-\frac{f_6}{6}+2\tilde{f_8}\biggr) e^{\nu\mu}\, \n_\nu\xi_\mu +\biggl(\frac{1}{12}+\frac{f_6}{6}+2\tilde{f_1}\biggr)e^{\mu\nu}\, \n_\nu \xi_\mu+\nonumber \\
&\,&  +\biggl(\frac{f_3}{12}+\frac{f_6}{6}+2\tilde{f_1}-2\tilde{f_8}\biggr)e^{\mu\nu}\,
\n^\alpha\Lambda_{[\alpha\mu\nu]} \Biggr\} \label{variacaoLams}
\end{eqnarray}
In order to obtain $\delta\lamsmzero=0$, we need to solve the equations below:
\begin{eqnarray}
-\frac{1}{24}+\frac{a_1}{2}+\frac{f_3}{6}+2\tilde{f_2}=0 \\
\frac{5}{24}-\frac{a_1}{2}-\frac{f_3}{6}-\frac{f_6}{6}+2\tilde{f_8}=0\\
\frac{1}{12}+\frac{f_6}{6}+2\tilde{f_1}=0\\
\frac{f_3}{12}+\frac{f_6}{6}+2\tilde{f_1}-2\tilde{f_8}=0
\end{eqnarray}
for which we have found the solution
\begin{eqnarray}
\tilde{f_1}=-\frac{1}{16}+\frac{a_1}{4}+\tilde{f_8}, \, \tilde{f_2}=-\frac{1}{16}-\frac{a_1}{4}-2\tilde{f_8}, \, f_3=1+24\, \tilde{f_8}, \, f_6=\frac{1}{4}-3a_1-12\, \tilde{f_8}\nonumber \\
\label{sol5282}
\end{eqnarray}
where $\tilde{f_8}$ is still arbitrary. The existence of such solution means that the Lagrangian $\lagmzero$ in
maximally symmetric spaces is symmetric under the full transformation given in (\ref{symla}).

\noindent By replacing the solution (\ref{sol5282}) back in $\lagmzero$ together with the fact that the space is
maximally symmetric, we reach the theory below:
\begin{eqnarray}
\lamsmzero &=&-\frac{1}{4}\n^\mu e^{\alpha\beta}\n_\mu e_{\alpha\beta}-\frac{1}{4}\n^\mu e^{\alpha\beta}\n_\mu e_{\beta\alpha}-\frac{1}{12}\n^\alpha e_{\alpha\beta}\n_\lambda e^{\lambda\beta}+\frac{1}{2}\n^\alpha e_{\alpha\beta}\n_\lambda e^{\beta\lambda}+\nonumber \\
&\,&+\frac{1}{4}\n^\alpha e_{\beta\alpha} \n_\lambda e^{\beta\lambda}+\frac{1}{6}\n^\mu \n_\mu e-\frac{1}{3}\n^\alpha e_{\alpha\beta}\n^\beta e-\frac{1}{24}\, R\,e^{\alpha\beta}\,e_{\alpha\beta}+\nonumber \\
&\,&+\biggl(\frac{1}{48}-\frac{a_1}{4}\biggr)\,R\,e^2+\biggl(-\frac{5}{48}+\frac{a_1}{4}\biggr)\,R
\,e^{\alpha\beta}\, e_{\beta\alpha}\label{lams2}
\end{eqnarray}
where the parameter $\tilde{f_8}$ has been naturally eliminated from the coefficients again. Therefore, we have
a model consistent with the description of massless spin-2 particles plus massless spin-0 particles propagating
in maximally symmetric spaces. Additionally we notice that (\ref{lams2}) is also consistent with the massless
limit of the massive $\lag$ model obtained in \cite{Fortes} for maximally symmetric spaces with
$\tilde{f_2}=-(a_1+1/4)/4$.
\section{Partially massless theories} \label{pmtheories}

In flat spaces, the particles are classified in massive or massless. On the other hand, on curved spaces (more
speciffically, in maximally symmetric spaces) there is another possible case where spin-2 particles can
propagate a number of degrees of freedom different from both massless and massive cases. The so called partially
massless theories which describe this kind of particles \cite{Hinter2,pm1,Hinter3} present a peculiarity:
although the mass is non-null, the theory has a scalar gauge invariance which is responsible for removing one of
the d.o.f. from the massive graviton. Let us see how this happens in the Fierz-Pauli theory.

In the subsection \ref{ssFPm0}, we have seen that in order to obtain the scalar constraint $h=0$ in the
Fierz-Pauli model, we need to demand the coefficient in (\ref{m2R}) to be non-null. However we have not analyzed
otherwise. Thus, let us consider that the coefficient of $h$ in (\ref{m2R}) is zero which leads us to:
\begin{eqnarray}
R=6m^2 \ .\label{relpm}
\end{eqnarray}
There is no scalar constraint $h=0$ anymore. Conversely, the theory acquires a scalar gauge symmetry:
\begin{eqnarray}
\delta h_{\mu\nu}=\n_\mu\n_\nu\phi +\frac{m^2}{2}g_{\mu\nu}\phi \label{trFp}
\end{eqnarray}
where $\phi$ is the gauge parameter.

On the other hand, the symmetry allows us to fix the gauge $h=0$. At this point, we have the same number of
d.o.f. of a massive spin-2 particle, which corresponds to 5 in $D=4$. However, even after choosing a specific
gauge, there is still a residual gauge invariance. More specifically, the theory remains invariant under a
subset of transformations (\ref{trFp}). If we perform the transformation (\ref{trFp}) again, the equations of
motion and the Fierz-Pauli constraints will remain unchanged. In order to verify it, let us first demand that
the trace $h$ remains null:
\begin{eqnarray}
h' = h+g^{\mu\nu}\, \delta h_{\mu\nu}= 0+(\square +2m^2)\phi
\end{eqnarray}
The new trace $h'$ will be zero if the parameter $\alpha$ satisfies the following equation
\begin{eqnarray}
\square\, \phi=-2m^2\phi \ . \label{solalpha1}
\end{eqnarray}
If we use (\ref{relpm}) and (\ref{solalpha1}), it is possible to verify  that the transverse condition $\n^\mu
h_{\mu\nu}=0$ will remain true and the equations of motion will not be modified.

Therefore, the residual gauge invariance given by (\ref{trFp}) and (\ref{solalpha1}) removes one more degree of
freedom from the theory. As a result we have 4 propagating d.o.f. instead of 5 which is called a partially
massless theory.

The partially massless theories have been studied at the linear level \cite{Hinter3,pm2,pm3,pm4} and there has
been a great effort to extend the studies to the non-linear level, despite the obstacles which have been raised
\cite{pm5,pm6}. They are of interest for the gravitational area since the equality (\ref{relpm}) implies a
direct relation between the graviton mass and the cosmological constant ($\lambda \propto R$). As we know, the
graviton mass, if it is non-null, would be very tiny, leading to an alternative to the cosmological constant
problem.

Let us see the partially massless theories associated to the $\lcg$ and $\lag$ models:
\begin{itemize}
\item $\lcg$

In \cite{Fortes} we have discussed in detail those massive models on curved spaces. The Lagrangian $\lc$ in
maximally symmetric spaces is the following: \bea
{\mathcal{L}}_{\mbox{\tiny{nFP}}}^g(c)&=&\sqrt{-g}\biggl[-\frac{1}{4}\n^\mu e^{\alpha\beta}\n_\mu e_{\alpha\beta}-\frac{1}{4}\n^\mu e^{\alpha\beta}\n_\mu e_{\beta\alpha}-\frac{1}{12}\n^\alpha e_{\alpha\beta}\n_\lambda e^{\lambda\beta}+\frac{1}{2}\n^\alpha e_{\alpha\beta}\n_\lambda e^{\beta\lambda}\nonumber \\
&\,&\hspace{1.2cm}+\frac{1}{4}\n^\alpha e_{\beta\alpha} \n_\lambda e^{\beta\lambda}+\frac{1}{6}\n^\mu \n_\mu e-\frac{1}{3}\n^\alpha e_{\alpha\beta}\n^\beta e-\frac{m^2}{2}(e_{\alpha\beta}e^{\beta\alpha}+c\, e^2)+\nonumber \\
&\,&\hspace{1.2cm}-\frac{1}{24}R
e^{\alpha\beta}e_{\alpha\beta}+\biggl(\tilde{d_2}+\frac{1}{12}\biggr)\,R\,e^2+\biggl(\tilde{d_8}-\frac{1}{24}\biggr)\,R\,e^{\alpha\beta}\,e_{\beta\alpha}\biggr]
\label{lcg2} \eea where $c$, $\tilde{d_2}$ and $\tilde{d_8}$ remain arbitrary. The manipulation of the equations
of motion obtained from $\lcg$ leads us to the necessary constraints in order to obtain the correct number of
degrees of freedom for a full massive theory, namely,
\begin{eqnarray}
e_{[\mu\nu]}&=&0\label{const1}\\
\n^\mu e_{\mu\nu}&=&0\label{const2}\\
e&=&0\label{const3}\ .
\end{eqnarray}

On the other hand, we have noticed that for a specific value of $R$, the theory acquires a scalar gauge
symmetry. More specifically, when
\begin{eqnarray}
(24\, \tilde{d_2}+1)R=12m^2c \label{Rpmlc}
\end{eqnarray}
and
\begin{eqnarray}
(24 c)\,\tilde{d_8}=24\tilde{d_2}-2c+1 \label{80}
\end{eqnarray}
the scalar symmetry
\begin{eqnarray}
\delta e_{\mu\nu}=\n_\mu \n_\nu \, \phi +\frac{R}{1+4c}\,  g_{\mu\nu}\, \phi\label{sympm2}
\end{eqnarray}
comes up. In this case, the constraints (\ref{const1}) and (\ref{const2}) remain true. However, the coefficient
of $e$ in the scalar constraint is identically null, excluding the possibility $e=0$.

Nevertheless, as happened in the Fierz-Pauli case, we can use the symmetry (\ref{sympm2}) in order to fix the
gauge $e=0$ leading us back to 5 d.o.f., which would be the correct counting for a massive spin-2 particle. But
there is still a residual gauge invariance. This can be seen if, after choosing the gauge $e=0$, we perform the
transformation (\ref{sympm2}) in the field $e_{\mu\nu}$ again. As a result, we obtain that all the equations and
constraints remain unchanged if
\begin{eqnarray}
\square \phi =-\frac{4R}{1+4c}\phi \ ,
\end{eqnarray}
where $R$ is given in (\ref{Rpmlc}). This choice removes one more degree of freedom and, consequently, we have
the so called partially massless theory for $\lc$ with 4 d.o.f. for a partially massless spin-2 particle.

\item $\lag$

Analogously, there is also a value of $R$ which gives rise to a scalar symmetry for the massive model $\lag$.
More specifically, if
\begin{eqnarray}
(6\tilde{f_8}+\frac{1}{2})R = 3\, m^2  \ ,\label{Rpm3}
\end{eqnarray}
the massive theory becomes invariant under the transformation
\begin{eqnarray}
\delta e_{\mu\nu}=\n_\mu\n_\nu\, \phi +\frac{R}{12}\, g_{\mu\nu}\, \phi \ .
\end{eqnarray}
Once again we can fix the gauge $e=0$. Even after fixing the gauge, we still have a residual gauge invariance.
Thus, we can remove one more degree of freedom from the theory by choosing $\phi$ in such a way that
\begin{eqnarray}
\square \phi = -\frac{R}{3} \phi
\end{eqnarray}
where $R$ is given in (\ref{Rpm3}).

Notice that $24 \, \tilde{d}_2 + 1 = 0 $ and $6\tilde{f}_8 + 1/2=0$ requires a fully massless theory ($m=0$),
see (\ref{80}) and (\ref{Rpm3}).

\end{itemize}

\section{Conclusion}

Lending continuity to the previous work \cite{Fortes} where we have studied the coupling of the new massive
models $\la$ to curved backgrounds, in the present work we have presented the analysis of the massless versions
of those models also on curved spaces.

In order to obtain the massless version of the $\la$ model coupled to a curved background, we have required the curved space versions of the corresponding flat space gauge symmetries. 
As a result, we have obtained a unique  model consistent with the description of a massless spin-2 particle
propagating in maximally symmetric spaces. It corresponds to the massless limit of a unique massive model from
\cite{Fortes}. Unfortunately, it was not possible to obtain the massless theories in more general background
spaces with the procedure used in our study, in contrast to the massless FP case which allows the propagation of
massless spin-2 particles on Einstein spaces \cite{Henn}. The key point is that instead of a ten component field
($h_{\mu\nu}=h_{\nu\mu}$) we have now a 16 components one ($e_{\mu\nu}\neq e_{\nu\mu}$) which requires a larger
symmetry, see (\ref{transflc}), than the linearized reparametrizations (\ref{eq13}) in order that we end up with
only two helicity modes ($\pm 2$) in the case of $\lc$ and an extra scalar mode in the $\la$ case. It turns out
that the tensor and vector symmetries in (\ref{transflc}) can hardly coexist on the curved space.

Additionally, partially massless theories have been found for the models $\lag$ consistently. We have been able
to find models with non vanishing mass with scalar gauge symmetries. This fact leads us to have theories which
describe, for some value of $R$, massive spin-2 particles with four degrees of freedom instead of the five
(2s+1) expected ones. Here we have gone beyond the initial studies of \cite{Fortes} and checked that the arising
scalar symmetry allows us to fix a gauge with residual symmetries consistent with four degrees of freedom. 

Finally, we are currently investigating  the addition of cosmological-like terms, $\Delta
\mathcal{L}=\sqrt{-g}\, [\Lambda_1 \, e_{\mu\nu}\,  e^{\mu\nu}+\Lambda_2 \, e_{\mu\nu}\, e^{\nu\mu}+\Lambda_3\,
e^2]$, to (\ref{La1gmzero}) and (\ref{lcg}) at $m=0$, altogether with singular terms on $\Lambda_j$ (linear in
curvatures) in the gauge transformations in order to achieve more general backgrounds.

\end{document}